# Load Mitigation and Power Tracking Control for Multi-Rotor Turbines

Horst Schulte[1], Urs Giger[2]

[1]University of Applied Sciences, HTW Berlin, Wilhelminenhofstr. 75a, 12459 Berlin, Germany

[2]GGS GmbH, Gotthardstraße 37, 6490 Andermatt, Switzerland

**Abstract:** A model-based feasible control strategy for multi-rotor systems is presented, pursuing two control objectives simultaneously: Mechanical loads on the main tower are to be mitigated, and an externally determined power change is to be followed to obtain fast power reference response in power systems. For this purpose, a scalable control strategy consisting of two levels is proposed: The first level consists of the decentralized control of each rotor unit. By using an LPV formalism, it is shown how the nonlinearities of the controlled system are considered in the design using decentralized wind speed observer of each rotor to improve the overall closed-loop performance. To mitigate the lateral loads on the multi-rotor main tower caused by asymmetric rotor thrust forces, a higher-level controller is introduced. Finally, the applicability of the controller structure is demonstrated by simulation studies.

## 1 Introduction

The concept of multi-rotors, i.e., replacing a single rotor with an equal-area number of three to seven rotors, was first explored in the 1980s until 1990s [Jam12]. One example is the Dutch manufacturer Lagerwey Wind, which has developed several multi-rotor systems. Here, promising approaches to important technical problems were solved, including rotor integration and yaw adjustment. Two and half decades later, in 2010, a multi-rotor wind turbine with seven rotors was tested at the NASA laboratory and provided encouraging results for future work [Jam14]. In addition, the manufacturer Vestas has developed a demonstrator of a multi-rotor system called R4-V29 with four commercial rotor units V29 [Lan19].

Manufacturers' developments have been driven by the advantage of the scaling law [Kal13, Jam18], according to which the total sum of the rotors and drive trains of the multi-rotor system can be significantly less in weight and cost compared to an equivalent single rotor system. Furthermore, the rotor interaction can cause an increase in the total power, as demonstrated on the R4-V29 prototype. The power curve measurements have shown that the rotor interaction of a multi-rotor turbine increases the power performance below the rated wind speed by 1.8 %, which can result in a 1.5% increase in annual energy [Lan19].

However, despite research and demonstrator developments by manufacturers over the past four decades, more systematic controller concepts for active load reduction on the mechanical multi-rotor structure need to be studied. The first investigations of the authors in [GIG21] have shown that due to the inclined flows, inhomogeneous wind fields, and asymmetrical operation in faulty cases large loads can occur in the tower structure. Also missing from previous studies are the turbine characteristics of power tracking, also known as active power control, which is necessary for grid-supporting and grid-forming operations.

Therefore, in this paper, firstly, a centralized multi-rotor control scheme is presented for compensating the bending moments in the tower structure due to the inhomogeneous wind fields. Second, decentralized controllers are used for fast power tracking of each rotor below the maximum producible power. The concept extends the previously presented active power tracking control (APC) scheme in [POE21] for single rotors.

## 2 Problem Formulation

### 2.1 System Overview and Control Requirements

In the investigations, a turbine with three rotors is examined. The characteristic feature compared to the Vestas turbine R4-V29 [Lan19] is that the tower and the rotor arms are designed as a steel lattice construction. The advantage is the flexible cost-saving delivery and space-saving installation, which can be carried out even in areas that are difficult to access. A study for a demonstrator to be installed in the Alpine region is shown in Figure 1. The tower itself is built on a natural hill, which explains the relatively low position of the rotors 2 and 3. The nominal turbine power is 3 x 30kW. A measurement container with variable ohmic inductive load is provided to investigate the power tracking capability.

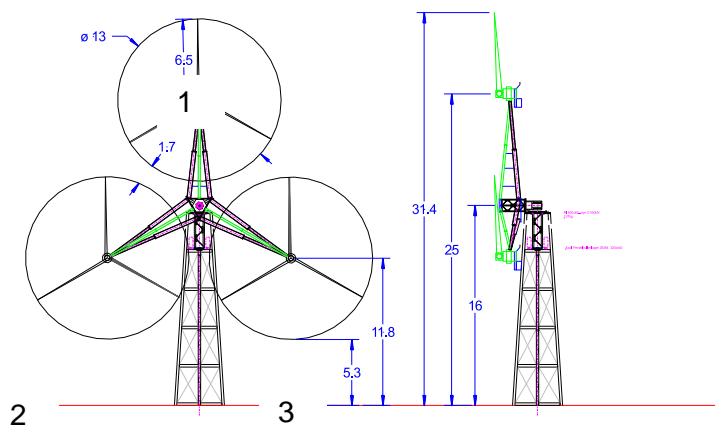

**Figure 1:** Multi-rotor prototype turbine with three 30kW rated power per rotor

For the design and development of the automatic control system, an overall concept is pursued. The rotors are operated with variable power that follows individual setpoints. Power dispatching is performed depending on the required load reduction in the tower structure. A scalable concept is required that can be easily adapted to a given multi-rotor topology in terms of the number of rotors and their arrangement. A structure is therefore being pursued in which the rotors are decentralized power-controlled, and a higher-level controller distributes the power distribution based on the load mitigation specifications.

In the next section, the turbine model is presented, based on which the simulation is created, and the reduced design model is derived.

## 2.2 Wind Turbine Modelling for Controller Validation

The model used for dynamics analysis and controller validation includes the aerodynamic thrust and torque per rotor, the structural dynamics model of the rotor arms and rotation of the tower around the main axis.

The rotor torque and trust force of each $i = 1,2,3$ turbine is defined by

$$T_{r,i} = \frac{1}{2} \rho \pi R^3 v_i^2 \, c_Q(\lambda, \beta), \qquad F_{T,i} = \frac{1}{2} \rho \pi R^2 v_i^2 \, c_T(\lambda, \beta), \tag{1}$$

where $T_{r,i}$ denotes the rotor torque and $F_{T,i}$ the trust force with the effective wind speed $v_i$ far in front of each rotor. The effect of aerodynamics on the rotor blades is described by stationary maps $c_Q(\lambda, \beta)$ and $c_T(\lambda, \beta)$ derived from the blade element theory [BUR21] illustrated in Figure 2 with the blade pitch angle $\beta$ and the tip speed ratio

$$\lambda = \frac{\omega_r R}{v} \,. \tag{2}$$

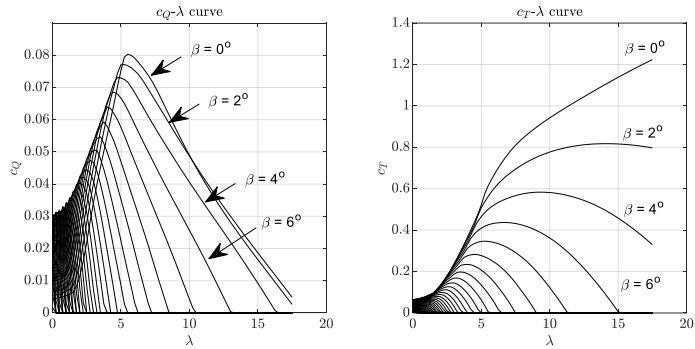

**Figure 2:** $c_Q - \lambda$ and $c_T - \lambda$ curves related to pitch angles $\beta \in \{0°, 2°, \dots, 88°, 90°\}$

For a realistic controller validation, a four degree of freedom (4 DOF) model of each *i*-th rotor-generator unit is used. The model is formulated for further analysis in a state space notation with separate representation of the linear (mechanical rigid body structure) and the nonlinear part with the aerodynamic torque and thrust force

$$\dot{x}_i = \begin{pmatrix} 0_{3\times 3} & A_{12} & 0_{3\times 1} \\ -M^{-1}K & -M^{-1}D & 0_{4\times 1} \\ 0_{1\times 3} & 0_{1\times 4} & -\frac{1}{\tau_\beta} \end{pmatrix} x_i + \begin{pmatrix} 0_{4\times 1} \\ \frac{1}{N\,m_B} F_T(x,v) \\ \frac{1}{J_r} T_r(x,v) \\ 0_{2\times 1} \end{pmatrix} + \begin{pmatrix} 0_{6\times 1} & 0_{6\times 1} \\ -\frac{1}{J_g} & 0 \\ 0 & \frac{1}{\tau_\beta} \end{pmatrix} u_i \tag{3}$$

with the state and input vector

$$x_i = \left( y_{a,i}, y_{b,i}, \Delta\theta_{S,i}, \dot{y}_{a,i}, \dot{y}_{b,i}, \omega_{r,i}, \omega_{g,i}, \beta_i \right)^T, \quad u_i = \left( T_{g,i}, \beta_{i,ref} \right)^T . \tag{4}$$

The state vector contains the generalized coordinates and their derivatives with $y_{a,i}$ as the fore-aft rotor arm deflection, $y_{b,i}$ the flap-wise tip deflection related to the body-fixed coordinates at the blade root, $\omega_{r,i}$ denotes the rotor speed and $\omega_{g,i}$ denotes the generator speed. The torsion angle related to the high-speed shaft is determined by

$$\Delta\theta_{S,i} = \theta_{r,i}\, n_g - \theta_{g,i}\;, \tag{5}$$

where $n_g$ is the gear box ratio. The matrices $M, K$ and $D$ denote the mass, stiffness and damping matrix and are given as follows

$$M = \begin{pmatrix} m_a + N\,m_b & N\,m_b & 0 & 0 \\ N\,m_b & N\,m_b & 0 & 0 \\ 0 & 0 & J_r & 0 \\ 0 & 0 & 0 & J_g \end{pmatrix}, \quad K = \begin{pmatrix} k_a & 0 & 0 & 0 \\ 0 & N\,k_b & 0 & 0 \\ 0 & 0 & k_S\,n_g^2 & -k_S\,n_g \\ 0 & 0 & -k_S\,n_g & k_S \end{pmatrix},$$

$$D = \begin{pmatrix} d_a & 0 & 0 & 0 \\ 0 & N\,d_b & 0 & 0 \\ 0 & 0 & d_S\,n_g^2 & -d_S\,n_g \\ 0 & 0 & -d_S\,n_g & d_S \end{pmatrix}. \tag{6}$$

Since the single rotors, the drive trains and rotor arm structure are identical, we assume that the aerodynamic curves shown in Figure 2 and the mechanical parameter values in (3), (5), and (6) are the same for all components of the multi-rotor turbine. These are denoted as such: $\tau_\beta$ is the time constant of the first-order model of the pitch drive of each rotor blade, $m_a$ is the effective mass of the rotor arm, $m_b$ is the effective mass of the rotor blade, $N$ is the number of blades, $J_r$ is the rotor inertia, $J_g$ is the generator inertia, $k_a$ is the effective rotor arm fore-aft stiffness parameter, $k_b$ is effective blade flap-wise stiffness parameter, $k_S$ is drive train stiffness parameter, $d_a$ is fore-aft rotor arm damping constant, $d_b$ is flap-wise blade damping constant, and $d_S$ drive train damping constant. For clarification of the equations of motion (3) and the corresponding generalized coordinates (4), schematic representations of one turbine unit with the rotor, mechanical structure and drive train are illustrated in Figure 3 and Figure 4.

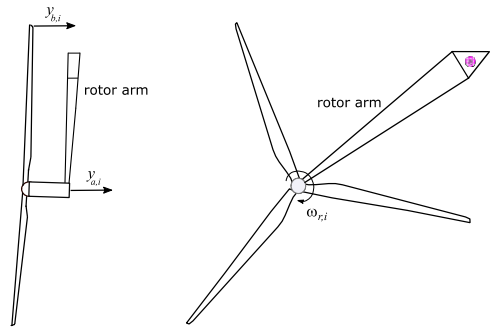

**Figure 3:** Schematic representation one turbine unit with rotor and mechanical structure with linkage to the overall multi-rotor turbine, see Figure 1

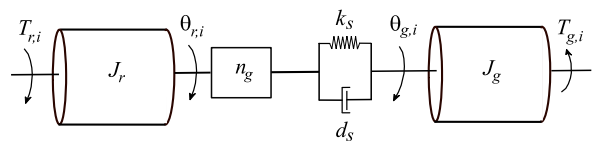

**Figure 4:** Schematic diagram of the drive train of each rotor-generator unit

Let us now consider the degree of freedom of the main tower of the multi-rotor system. Preliminary investigations have shown that asymmetric thrust forces on the rotors result in large bending moments in the tower base. This can have several causes: In normal operation, it can be caused by a non-homogeneous wind field distribution, or due to an inclined flow with a wake effect. Furthermore, faulty sensors for position and speed measurement and actuator faults in the pitch drive and generator can lead to highly asymmetrical loads. However, the faulty case is not the subject of this study and will be investigated elsewhere in the development of Fault tolerant control (FTC) strategies. We consider normal operation, where an overall central control strategy significantly reduces the torsion of the tower in the main axis and thus the load in the tower base.

Let us now look at a simplified top view of the multi-rotor system (Figure 1) without rotor 1 illustrated in Figure 5.

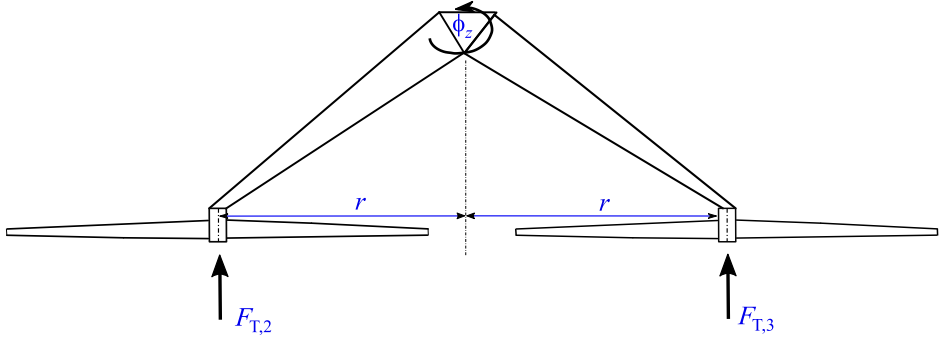

**Figure 5:** Top view of the multi-rotor system from Figure 1 without rotor 1

Using the defined coordinates and thrust forces at the individual rotor 2 and 3, the following equation of motion is obtained:

$$J_z \ddot{\phi}_z = -k_z \phi_z - d_z \dot{\phi}_z + r \left( F_{T,3} - F_{T,2} \right) \tag{7}$$

the torsion angle $\phi_z$, where $J_z$ denotes the inertia of the multi-rotor system, $k_z$ denotes the main tower stiffness parameter and $d_z$ the damping parameter. All effective parameters are related to the z-axis.

## 3  Controller Structure and Model-based Design

The control system consists of a decentralized power tracking control of each rotor unit and a higher-level control for load reduction on the overall system. Here we focus on mechanical loads that can lead to torsional moments at the tower's base due to inhomogeneous wind fields

### 3.1  Decentralized Control for Power Tracking

The LPV decentralized controller scheme for the active power control of each rotor-generator unit above the design wind speed is shown in Figure 6. A further structure could be given for the area below the design wind speed but cannot be shown here due to the limited amount of space available.

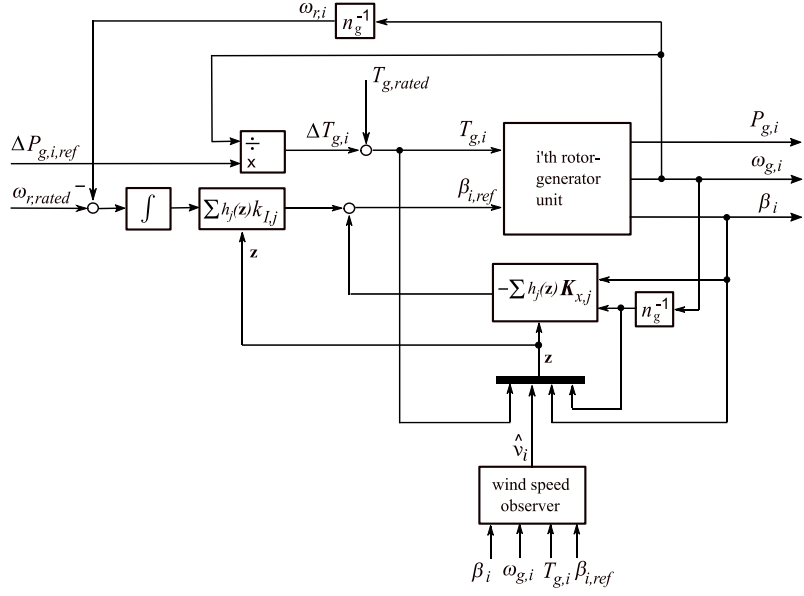

**Figure 6:** Local LPV power tracking controller of *i*'th rotor-generator unit

The time-varying reference variable of the control is the required power change, which is denoted by $\Delta P_{g,i,ref}$. A fixed operating point, on the other hand, is the common rated rotor speed $\omega_{r,rated}$. Thus, one achieves a rapid change of the power, with the variation of the generator torque $T_{g,i}$. The resulting torque imbalance is compensated by the setpoint controlled tracking of the pitch angle $\beta_i$. The controller coefficients $K_{x,j}$ and $k_{I,j}$ (see Figure 6) are calculated model-based from a 1-DOF nonlinear turbine model in Takagi-Sugeno form (like a quasi-LPV system) augmented by first order pitch dynamics using LMI (Linear Matrix Inequality) design criteria [POE21]. The control-oriented model is obtained by model reduction of the 4-DOF state-space model (4):

$$\dot{x} = \sum_{i=1}^{N_r=4} h_i(z)\, A_i\, x + B\, u, \qquad y = C\, x, \tag{8}$$

with the premise variable $z = (\omega_{r,i}, \beta_i, v_i, T_{g,i})^T$, the state vector $x = (\omega_{r,i}, \beta_i)^T$ and input $\beta_{i,ref}$. Note that in the LPV framework the $z$ is equivalent to the parameter vector $\theta$. The variable $v_i$ denotes the undisturbed effective wind speed in front of the i'th rotor. Here it is reconstructed online with the wind speed observer from Figure 6 from the measured states and the inputs.

## 3.2 Higher-level Control for Load Mitigation

The central control system consists of the PI controller law

$$u_c = k_{p,c}\, \dot{\phi}_z + k_{I,c} \int_0^t \dot{\phi}_z\, dt \tag{9}$$

and an adaptive dispatcher to determine the changes of the reference powers $\Delta P_{g,i,ref}$, $i = 2,3$ for the local LPV controllers of rotor 2 and 3. With $u_c$, a participation factor is determined for each individual turbine. If $u_c = 0$, the demanded total power change $\Delta P_{g,2\&3,ref}$ of unit 2 and 3 is allocated equally to the turbines. If $u_c \neq 0$ the power is distributed to the turbines in such a way that asymmetric dynamic forces $F_{ST,i} = \frac{1}{2}\, \rho\, \pi\, R^2\, v_i^2$ on the rotors due to a non-homogeneous wind field ($v_i \neq v_j$) are compensated via different $c_T$ adjustment (see Figure 2) so that the torsion angle characterized by (7) is decreased and thus the loads on the tower are reduced. In detail, the reference values for lower-level decentralized controllers of unit 2 and 3 are calculated as follows

$$\Delta P_{g,2,ref} = \left(\frac{1}{2} + u_c\right)\Delta P_{g,2\&3,ref} \; , \quad \Delta P_{g,3,ref} = \left(\frac{1}{2} - u_c\right)\Delta P_{g,2\&3,ref} \; . \tag{10}$$

In the next section, the performance of the approach for a three-rotor system with a rated power of 3x5MW is presented. It should be noted that the method presented here is not limited to three rotors and can therefore be extended to four, five or more rotor units.

## 4   Simulation Results

The control concept for power tracking with load reduction using a turbulent wind field was investigated. Up to time t=50 s, the wind field is still uniform. After that, the wind speeds at rotors two and three differ by 1 m/s, see the upper diagram of Figure 7. Furthermore, the total power reference is gradually reduced, whereby up to t=70s the power is still dispatched equally to the rotors (lower diagram). A metric for the load at the base of the tower is the torsion angle $\phi_z$ at the star point of the tower (see Figure 1). It can be clearly seen that due to the no longer homogeneous wind field from t=50s, the loads increase immediately. By switching on the load-mitigation controller at t=70s, which means a difference in the power generation of rotor unit 2 and 3, the load at the base of the tower is significantly reduced. A closer look shows the shift in the generator power proportions from t = 70s, although the sum of them does not change.

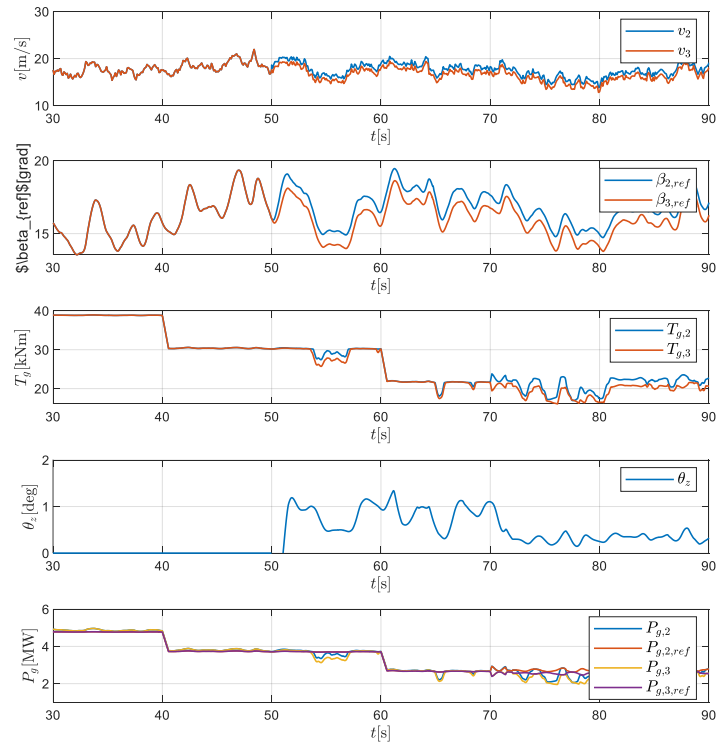

**Figure 7:** Simulation study without (t= 50 to 70 s) and with load mitigation (t>70s).

# 5  Conclusion

A control concept for a multi-rotor turbine was presented. The focus was on the presentation of the basic structure, the simulation models used and the clarification of the effect of the control laws. Not described in detail was the exact model-based design procedure with LMI criteria. The interested reader will find a detailed description of the design process in [POE21], but not yet the adaptive distributed power tracking exploited for load-mitigation This has been presented for the first time in this work.

# 6  Bibliography

[Jam12]   Jamieson, Peter and Branney, Michael: Multi-Rotors: A Solution to 20 MW and Beyond? In: Energy Procedia, 24: 52–59, 2012

[Jam14]   Jamieson, Peter; Branney, Michael: Structural Considerations of a 20 MW Multi-Rotor Wind Energy System. In: IOP Conf. Series: Journal of Physics: Conf. Series 012013, 2014

[Kal13]   A. Kale, Sandip and Sapali, S. N.: A Review of Multi-Rotor Wind Turbine Systems. In: Journal of Sustainable Manufacturing and Renewable Energy 2(1/2):60–68, 2013

[Jam18]   Jamieson, Peter: Innovation in Wind Turbine Design, 2nd Edition. John Wiley & Sons. 2018.

[LAN19]   van der Laan, Maarten Paul; Andersen; Juhl, Søren; García, Néstor Ramos; Angelou, Nikolas, Pirrung, Georg Raimund; Ott Søren; Sjöholm, Mikael; Sørensen, Kim Hylling; Neto, Julio Xavier Vianna; Kelly, Mark; Mikkelsen, Torben Krogh, and Larsen, Gunner Christian: Power curve and wake analyses of the Vestas multi-rotor demonstrator. In: Wind Energy Sciences, 4, 251–271, 2019 https://doi.org/10.5194/wes-4-251-2019

[GIG21]   Giger, Urs; Kleinhansl, Stefan; Schulte, Horst: Design Study of Multi-Rotor and Multi-Generator Wind Turbine with Lattice Tower - A Mechatronic Approach. In: Applied Sciences 11(22), 2021, https://doi.org/10.3390/app112211043

[POE21]   Pöschke, Florian, Schulte, Horst: Model-based control of wind turbines for active power control - A unified sector-nonlinearity approach based on Takagi-Sugeno modeling. In: at – Automatisieungstechnik 69, 10, 820-835, 2021, https://doi.org/10.1515/auto-2021-0047

[BUR21]   Burton, Tony L.; Jenkins, Nick; Bossanyi, Ervin; Sharpe, David; Graham, Michael: Wind Energy Handbook, 3rd Edition, April 2021